\documentclass[lettersize,journal]{IEEEtran}
\usepackage{amsmath,amsfonts}
\usepackage{algorithmic}
\usepackage{algorithm}
\usepackage{array}
\usepackage{textcomp}
\usepackage{stfloats}
\usepackage{url}
\usepackage{verbatim}
\usepackage{graphicx}
\usepackage{xcolor}
\usepackage{hyperref}
\usepackage{algorithm}
\usepackage{algorithmic}

\usepackage{graphicx}
\usepackage{subcaption}
\usepackage[numbers]{natbib}
\usepackage{orcidlink}
\newcommand{\result}[1]{}

\definecolor{myred}{cmyk}{0, 0.7808, 0.4429, 0.1412}

\newcommand{\done}[1]{}

\usepackage{pifont}

\usepackage{xspace}
\newcommand{\etal}{\textit{et al.}~}
\newcommand{\eg}{\textit{e.g.,}\xspace}
\newcommand{\ie}{\textit{i.e.,}~}

\newcommand{\one}{({\em i})\xspace}
\newcommand{\two}{({\em ii})\xspace}
\newcommand{\three}{({\em iii})\xspace}

\makeatletter
\newcommand{\paragraphNoDot}[1]{\vspace*{0.03in}\noindent{\bf #1}\hspace{0.25ex \@plus1ex \@minus.2ex}}
\makeatother

\usepackage{arydshln}

\begin{document}

\title{Deconstructing BLE Multi-hop: a Model-based Approach to Quantifying the Challenges}

\author{%
Bozheng~Pang\,\orcidlink{0000-0001-6709-8813},
Jos\'e~Alamos\,\orcidlink{0000-0001-5681-2591},
Thomas~C.~Schmidt\,\orcidlink{0000-0002-0956-7885},
and Matthias~W\"ahlisch\,\orcidlink{0000-0002-3825-2807}%
\thanks{B. Pang and M. W\"ahlisch are with the Faculty of Computer Science, TU Dresden, Dresden, Germany. E-mail: \{bozheng.pang, m.waehlisch\}@tu-dresden.de.}%
\thanks{M. W\"ahlisch is also with Barkhausen Institut, Dresden, Germany.}%
\thanks{J. Alamos and T. C. Schmidt are with the Department of Computer Science, HAW Hamburg, Hamburg, Germany. E-mail: \{jose.alamos, t.schmidt\}@haw-hamburg.de.}%
}

\markboth{Journal of \LaTeX\ Class Files,~Vol.~14, No.~8, August~2021}%
{Shell \MakeLowercase{\textit{et al.}}: A Sample Article Using IEEEtran.cls for IEEE Journals}

\maketitle

\begin{abstract}
\label{Abstract}
Bluetooth Low Energy (BLE) was originally designed for point-to-point communication, but BLE multi-hop networks have attracted academic and industrial interest.
Multi-hop communication, however, introduces desynchronized nodes to the network due to the difference in individual clocks.
In this paper, we study the impact of clock drift on a BLE multi-hop network.
The aim is to find out if a BLE multi-hop network can be properly set up, why, and how.
For this goal, we develop a mathematical model to quantify the collision probability among BLE connections managed by a single device.
We provide a simplified version of the model for constrained deployment.
The impact of clock drift and our mathematical model are validated through practical experiments.
Afterward, we find that a BLE multi-hop network cannot eliminate the impact of clock drift in most cases.
Moreover, from a long-term point of view, the collision between BLE connections occurs with a fixed frequency.
We further analyze the limitation of using BLE to form multi-hop networks, and propose possible solutions, such as topology limitation, changes to BLE stack or hardware.
This paper is a clear and quantitative problem statement on the limitations of using BLE in multi-hop networks.
\end{abstract}


\section{Introduction}
\label{Introductiion}
Bluetooth Low Energy (BLE)~\cite{bluetooth_sig_bluetooth_2024} is a wireless protocol specifically designed for point-to-point communication.
Since industry increasingly prioritizes sustainability and connectivity of their machines and products, BLE plays a key role in more and more applications and scenarios~\cite{koulouras_evolution_2025}.
Nowadays, it is a default technology implemented in almost every laptop and smartphone~\cite{koulouras_evolution_2025}.
Thus, it is also a promising network access in Internet of Things (IoT) deployments.

Single-hop communication is easy to deploy and manage but lacks flexibility and limits the communication range.
Many BLE deployments require flexible topologies~\cite{leonardi_multi-hop_2018}, allowing a single node to multiplex resources between multiple directly connected devices.
Time-triggered protocols such as BLE, however, are specifically challenged in multi-hop setups since they require multiple nodes to adhere to a strict schedule in parallel, which determines when to send and when to listen on which channel and time slots.
Such strict timing is hard to guarantee on commodity IoT devices due to hardware imperfections, which easily lead to clock drifts~\cite{bankov_clock_2018,petersen_mind_2021,asgarian_bluesync_2022}.

In this paper, we start from the observation that support of multi-hop BLE networks is important but the impact of clock drifts on the performance is not fully understood yet.
We systematically uncover the limitations of BLE when establishing a multi-hop network and clock drift occurs.
We consider this paper a clear and quantitative problem statement as the basis for later research on BLE multi-hop networks.
Our results show that a BLE multi-hop network cannot eliminate the impact of clock drift in most cases.
In all topologies, except the star topology where a BLE central is connected to multiple BLE peripherals, the clock drift will disturb communication performance.
We find that the collision of events caused by clock drift has a fixed frequency of occurrence, resulting in packet losses in long-term deployments.
In scenarios with volatile topologies and short communication setups, some fine-tuning on BLE connection parameters can help optimize the network.
In detail, our contributions are the following:

\begin{enumerate}
	\item
	We validate the clock drift phenomenon through practical experiments.
	\item
	We introduce a mathematical model to evaluate the performance of BLE multi-hop networks.
	This model allows us to identify systematical limits because of clock drifts.
	\item
	We provide a simplification of our model, which reduces computational complexity without introducing significant errors.
	\item
	We validate our models through practical experiments.
\end{enumerate}

Prior work~\cite{petersen_mind_2021} empirically identified clock drift as a major reason for performance drawbacks in BLE multi-hop networks.
To counter overlapping time slots, the authors proposed artificial acceleration of drifts.
This approach mitigated breaking connections in a specific deployment.
Our work fills the gap by providing a theoretic model that allows to reveal the limits of multi-hop networks given local clock drifts.
This model is independent of specific deployments and allows to clearly identify when multi-hop BLE networks are suitable and when not.

The remainder of this paper is structured as follows.
In \autoref{Background}, we present background on BLE and the core challenge introduced by clock drifts on BLE multi-hop networks.
We derive our mathematical model in \autoref{Mathematical Model}, and discuss a simplification for practical use in \autoref{Model Approximation}.
In \autoref{Analysis Experiment}, we present our experiments to analyze the impact of clock drifts based on real hardware deployed in different setups.
In \autoref{Verification Experiment}, we validate our mathematical model and its simplification by comparing with simulations.
Related work is discussed in \autoref{Related Works}.
We address limitations and potentials of using BLE in multi-hop networks based on our findings in \autoref{Limitation_Analysis}.
We conclude with an outlook in \autoref{Conclusions}.

\section{Background}
\label{Background}
\subsection{Basic Knowledge}
The BLE standard~\cite{bluetooth_sig_bluetooth_2023} defines communication modes such as connectionless and connection-oriented.
According to 6LoBLE and the Internet Service Support Profile~\cite{gomez_ipv6_2021}, IP data should be transferred using the connection-oriented mode.

A simplest BLE connection involves two devices: a central and a peripheral~\cite{bluetooth_sig_bluetooth_2023}.
Since BLE 4.2~\cite{bluetooth_sig_bluetooth_2014}, a device can hold both roles simultaneously.
The timeline of a BLE connection is divided into many continuous connection intervals.
The value of a connection interval can range from 7.5~ms to 4~s~\cite{bluetooth_sig_bluetooth_2023}.
Each interval begins with a central packet, followed by a peripheral one.
This forms a data transaction~\cite{pang_modeling_2024}.
Multiple transactions can occur in each connection interval, and a connection event is made up of all the transactions inside the same interval~\cite{bluetooth_sig_bluetooth_2023}.

To ensure reliability, BLE employs adaptive frequency hopping (AFH)~\cite{bluetooth_sig_bluetooth_2023}.
It divides the 2.4~GHz frequency band into 37 data channels and 3 advertising channels~\cite{bluetooth_sig_bluetooth_2023}.
A BLE connection hops across the data channels pseudo-randomly to avoid possible interference~\cite{bluetooth_sig_bluetooth_2023}.
The hopping occurs at the beginning of each connection interval.
Then the connection stays on the channel for the entire interval.
This starting point is called an anchor point.

To maintain a BLE connection, both central and peripheral must follow a shared schedule strictly.
At each anchor point, they should hop to the same channel, and the peripheral should be ready to receive while the central sending its packet.
Due to clock drift~\cite{petersen_mind_2021}, the BLE standard~\cite{bluetooth_sig_bluetooth_2023} sets the schedule of the central as the benchmark, and the peripheral calibrates its schedule accordingly.
The peripheral estimates the anchor point and opens a window around it.
By listening throughout this window, it can find the central and calibrate its schedule.
By implementing all the details above, a long-term connection between two BLE devices can be kept~\cite{bluetooth_sig_bluetooth_2023}.

\subsection{Challenge in Multi-hop}
A BLE multi-hop network involves more than two BLE devices.
If there is only one central, all peripherals will follow the central schedule.
This should allow proper time management.
However, with multiple centrals, multiple schedules exist.
This will lead to potential collisions for devices involved in multiple connections.

\begin{figure}
	\centering
	\includegraphics[width=0.7\linewidth]{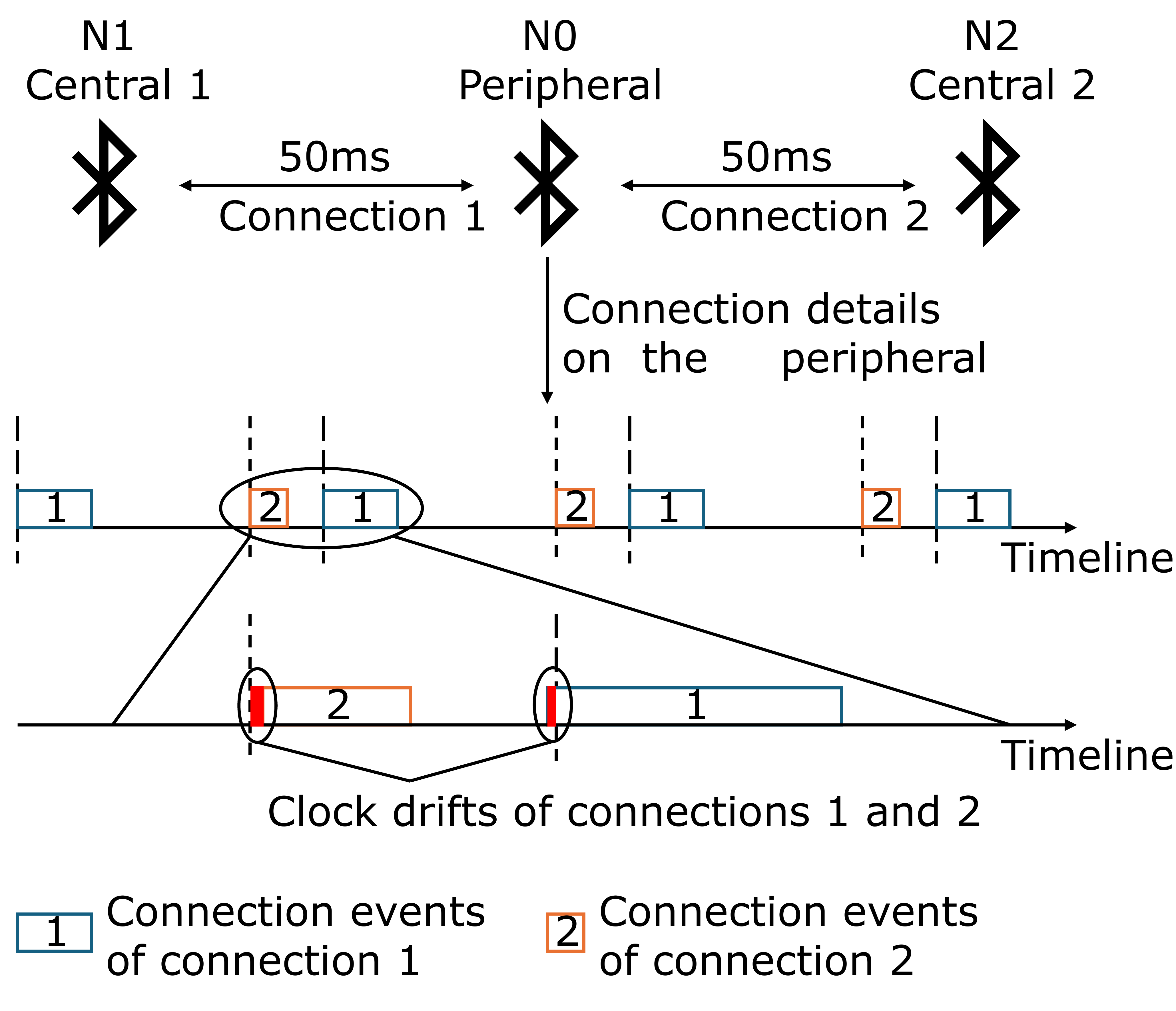}
	\caption[]{A simple BLE network containing three nodes. The peripheral (N0) calibrates its two connection schedules to the two centrals (N1, N2), thus a relative drift between the two connection schedules exists on the peripheral.}
	\label{fig:fig01}
\end{figure}

\autoref{fig:fig01} shows a simple network with two centrals (N1, N2) and one peripheral (N0).
A typical setup is that their connection intervals are default, thus the same.
Since N0 must calibrate itself to two central schedules, differences between them cause the anchor points to gradually drift over time.
This drift can accumulate and lead to connection event collisions.
Understanding this issue is important for deploying BLE multi-hop networks.
This phenomenon was first reported in~\cite{petersen_mind_2021} and further validated in \autoref{Analysis Experiment}.

\section{Mathematical Model}
\label{Mathematical Model}
In this section, we develop a mathematical model to analyze and quantify the collision probability across multiple connections managed by a single BLE device.
To simplify the derivation, we start with a simple topology (\autoref{fig:fig01}), then extend the results to more complex cases.
We also introduce a graphical method to ease the understanding of the mathematical model.

In case some BLE designers and developers might want to implement our model on constrained IoT devices as a tool to monitor and manipulate their BLE networks, we offer a simplified version of the model in \autoref{Model Approximation}.
This is to avoid too complex and impractical implementation on constrained IoT devices, such as some existing mathematical approaches~\cite{self_intercept_1985, clarkson_numbertheoretic_1996}.

\begin{figure}%
	\centering
	\includegraphics[width=0.7\linewidth]{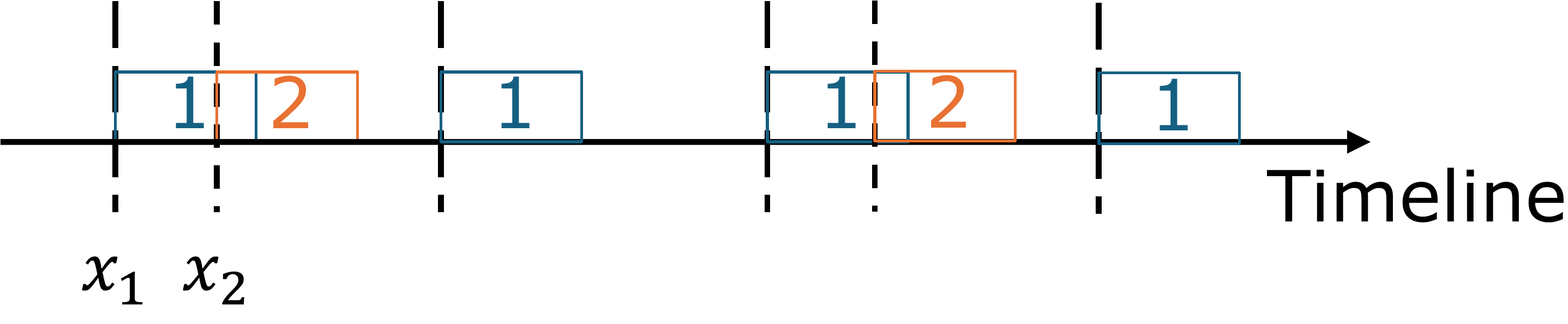}
	\caption{Given connection~1 and 2, it is important to calculate the collision probabilities for both connections separately. For both connections, two collisions occur, but the collision probability is 50\% for connection~1 and 100\% for connection 2. $x_1$ and $x_2$ are the anchor points of the corresponding connection intervals.}
	\label{fig:fig02}
\end{figure}

It is worth noting that we calculate the collision probabilities of two connections separately since two connections may lead to the same number of collisions but to different probabilities, depending on the connection interval values~(see \autoref{fig:fig02}).

\subsection{Overlap Probability}
\begin{figure}
	\centering
	\includegraphics[width=0.6\linewidth]{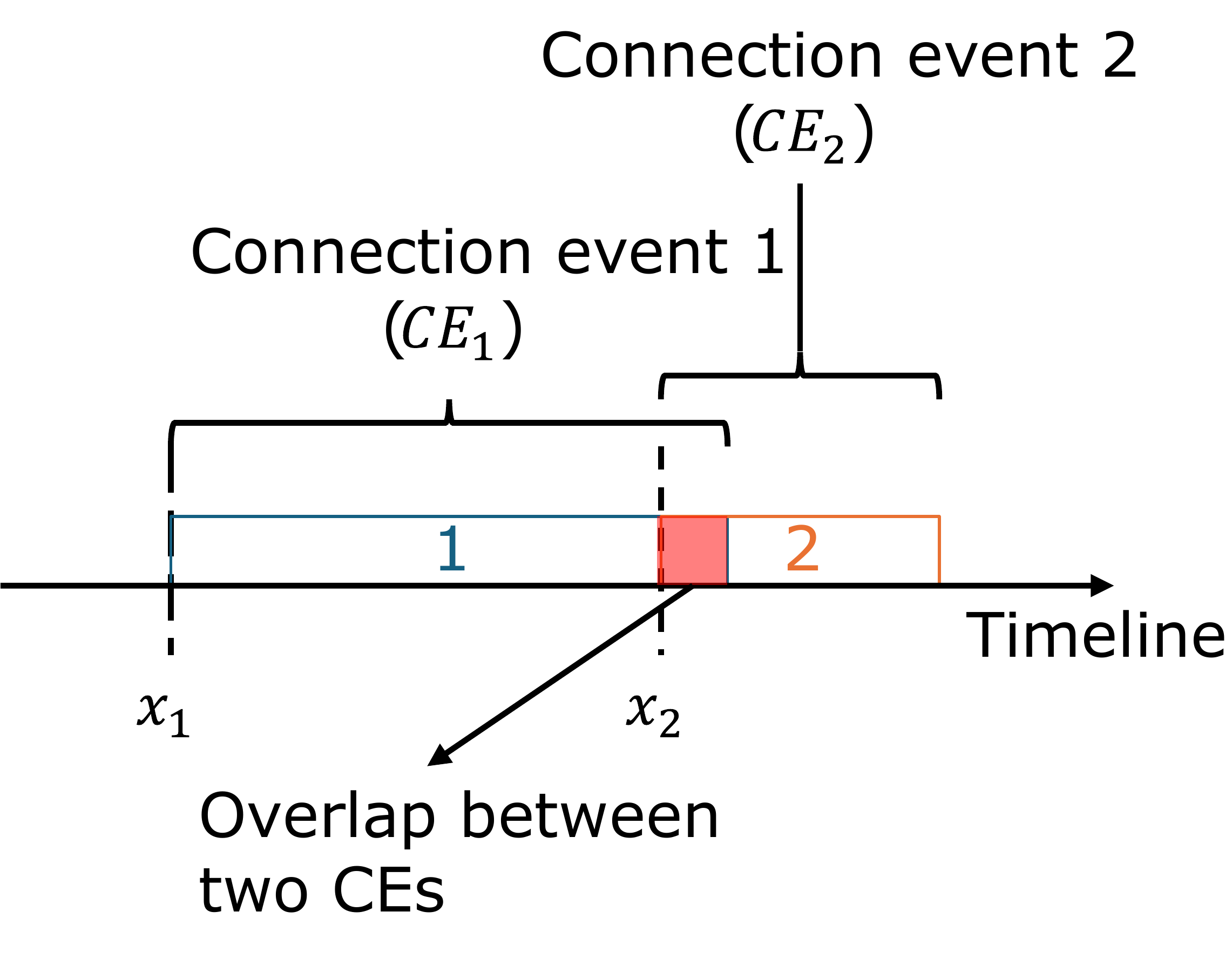}
	\caption{Overlap between two BLE connections. An overlap occurs once two connection events exist simultaneously.}
	\label{fig:fig03}
\end{figure}

In this paper, we define the collision probability as the ratio between the number of overlapped connection events and the number of total connection events.
The graphical expression of an overlap between two random connection events is shown in \autoref{fig:fig03}.
Whenever a connection event ends later than the anchor point of the other connection event, there is an overlap.

Let $x_1$ and $x_2$ denote the anchor points of two connection events, and $CE_1$ and $CE_2$ represent the connection event lengths in time.
Then, the following conditions lead to an overlap:

\begin{equation}
	\label{equation01}
	\small
	x_2 < x_1 + CE_1
	\textnormal{\quad and \quad} x_1 < x_2 + CE_2
\end{equation}

By organizing the two conditions in \autoref{equation01}, we find the following relation among $x_1$, $x_2$, $CE_1$, and $CE_2$.

\begin{equation}
	\label{equation02}
	\small
	-CE_1 < x_1 - x_2 < CE_2  
\end{equation}

In practice, there can be a huge difference between the starting time of two connections, \eg two hours.
However, both anchor points $x_1$ and $x_2$ are limited by the ranges $[0, CI_1]$ and $[0, CI_2]$, respectively.
This is because the periodicity of anchor points makes it meaningless to extend $x_1$ and $x_2$ over the given ranges.
By going over the given ranges, we can cover all possible combinations between $x_1$ and $x_2$.
For example, the combination of $x_1 = 0$ and $x_2 = 0$ repeats the combination of $x_1 = CI_1$ and $x_2 = 0$.
This is because the anchor point before $x_1 = CI_1$ on connection~1 is $x_1 - CI_1 = 0$.

Combining \autoref{equation02} with the boundaries of $x_1$ and $x_2$, we can list all possible overlap cases.
However, since $x_1$ and $x_2$ are continuous variables, the list of all possible overlap cases turns out to be infinite~\cite{anton_calculus_2021}.
To quantify the amount of the possible overlap cases~($A_1$), we derive the following equation.

\begin{equation}
	\label{equation03}
	\small
	A_1 
	= \int_0^{CI_1} \int_0^{CI_2} 
	1\{-CE_1 < x_1 - x_2 < CE_2\} 
	\,dx_1\,dx_2
\end{equation}

\autoref{equation03} involves a conditional equation and a double integral.
The conditional equation is $1\{-CE_1 < x_1 - x_2 < CE_2\}$.
When $x_1 - x_2$ is within $-CE_1$ and $CE_2$, the output is 1, otherwise it is 0~\cite{anton_calculus_2021}.
To find all possible overlap cases, an integration is applied to the conditional equation.
Since $x_1$ and $x_2$ are the only two variables inside, we use a double integral.
This integral achieves quantification of the possible overlap cases when varying $x_1$ and $x_2$.

In BLE, connection intervals~($CI_1$, $CI_2$) typically remain constant, thus can be considered static variables.
Connection event lengths~($CE_1$, $CE_2$) depend on the application but also can be considered static variables, \eg the data transfer in a BLE multi-hop network connected to the Internet.
During idle periods, short connection events with empty packets are used to keep the connections alive.
During active periods, long events are used to quickly transfer the data.
Overall, both connection intervals and event lengths can be considered static.

For the convenience of explaining and understanding, we derive the overlap probability for the connection with a smaller connection interval first. 
After that, the collision probability for the other connection (with a larger connection interval) can be easily found, assuming $CI_1 < CI_2$.
Note that we explicitly exclude the case of $CI_1 = CI_2$ since it obviously causes connection shading as empirically shown in prior work~\cite{petersen_mind_2021}.

Based on the total amount of overlap cases $A_1$ derived in Equation \ref{equation03}, we can calculate the overlap probability.
It is equal to the ratio between the total amount of overlap cases and all the possible combinations of $x_1$ and $x_2$.
For connection~1, the overlap probability with connection~2 is:

\begin{equation}
	\label{equation04}
	\footnotesize
	\begin{aligned}
		P_{12} 
		& = \frac{A_1}{CI_1 \cdot CI_2} \\
		& = \frac{1}{CI_1 \cdot CI_2} 
		\int_0^{CI_1} \int_0^{CI_2} 
		1\{-CE_1 < x_1 - x_2 < CE_2\} 
		\,dx_1\,dx_2
	\end{aligned}
\end{equation}

The product of $CI_1$ and $CI_2$ quantifies all the possible combinations of $x_1$ and $x_2$.
\autoref{equation04} calculates how likely two connection events overlap for connection~1 when randomizing their anchor points $x_1$ and $x_2$.

\subsection{Collision Probability}
To develop the collision probability based on the overlap probability, we introduce the following graphic method.
We believe that our graphic method can better explain the collision probability.
It also gives an intuitive understanding to the later contents, \eg why a multi-hop BLE network cannot avoid the impact of clock drift.

\begin{figure}
	\centering
	\includegraphics[width=1\linewidth]{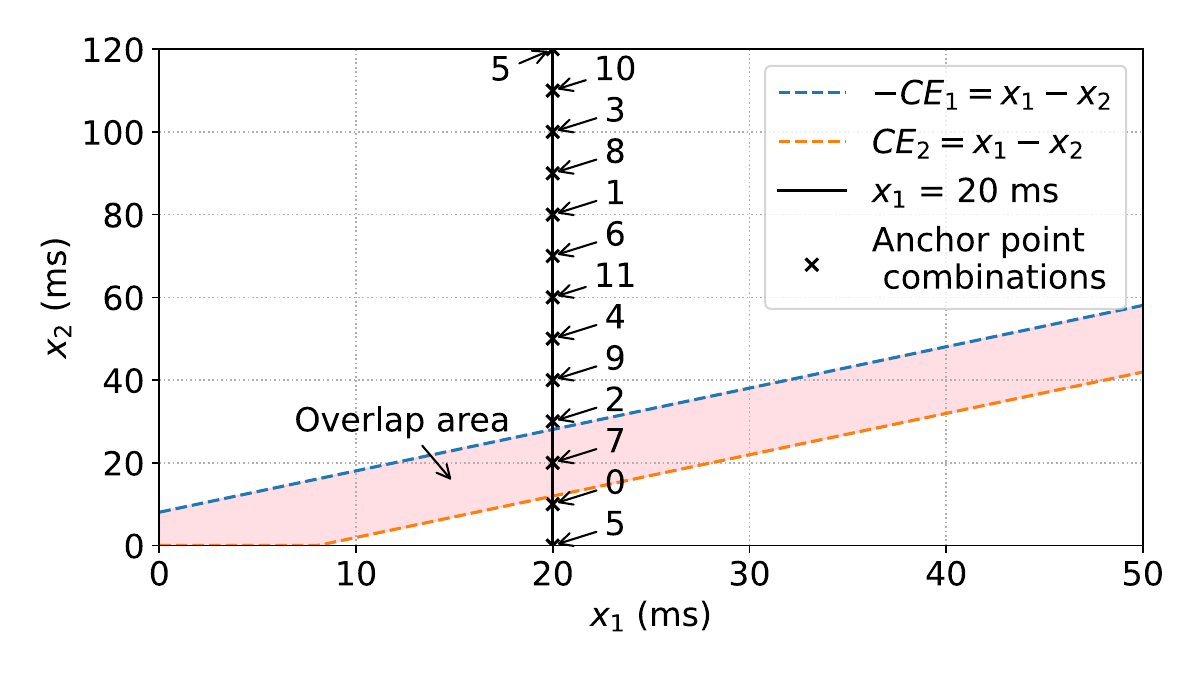}
	\caption{Graphical representation and explanation of \autoref{equation04}. A random point on the figure represents a combination of two anchor points of two connections. The location of the point tells if there is an overlap between the connection events following the two anchor points. Note that this figure represents the case for the connection with a smaller interval value.}
	\label{fig:fig04}
\end{figure}

\autoref{fig:fig04} graphically represents \autoref{equation04}.
$x_1$ and $x_2$ are plotted on the $x$- and $y$-axes, with the boundaries of $[0, CI_1]$ and $[0, CI_2]$, respectively.
In our example, we assign $CI_1 = 50$~ms and $CI_2 = 120$~ms.
The dashed blue and orange lines represent \autoref{equation02}, and the red area highlights the overlap.
A random point on the figure corresponds to a pair of $x_1$ and~$x_2$.
Depending on whether the point is located in the red area or not, it tells whether an overlap occurs.
Using the Monte Carlo method~(\ie throwing numerous random points onto the figure), one can estimate the overlap probability by dividing the number of points in the red area by the total points~\cite{barbu_monte_2020}.
Alternatively, we can obtain an accurate result by dividing the size of the red area by the total area ($50 \times 120 = 6000$).

Now, we assume a random point on the figure, \eg the point~0~($x_1 = 20$ and $x_2 = 10$).
This means that connection~1 has an anchor point at 20~ms, while connection~2 at 10~ms.
It is worth noting the difference between the two connection intervals, \ie 70~ms~($CI_2 - CI_1 = 120 - 50$).

To calculate the collision probability for connection~1, we must go over all the connection events of connection~1 to find all the overlaps between connection~1 and connection~2.
According to \autoref{equation02} and \autoref{fig:fig04}, the two initial anchor points, 20~ms and 10~ms, do not cause any overlap.
By adding the corresponding connection interval values to the two initial anchor points, the next two anchor points can be calculated.
They are 70~ms~($20\mathrm{~ms} + 50\mathrm{~ms} $) and 130~ms~($10\mathrm{~ms}  + 120\mathrm{~ms} $).
\autoref{fig:fig04} cannot show these two anchor points since they are out of the plot boundaries of the figure.
To draw them in \autoref{fig:fig04}, we keep the time difference~($130\mathrm{~ms} - 70\mathrm{~ms} = 60\mathrm{~ms}$) between them the same, and move these two anchor points forward on the timeline to 20~ms~($70\mathrm{~ms} - 50\mathrm{~ms}$) and 80~ms~($130\mathrm{~ms} - 50\mathrm{~ms}$).
This way, point~1 in \autoref{fig:fig04} represents them.
Comparing point~1 with point~0, the $x_1$ value remains the same, but the $x_2$ value changes by 70~ms~(the difference between the two connection intervals).
By adding another connection interval value, we can find the next anchor point of connection~1, which is 120~ms~($20\mathrm{~ms} + 50\mathrm{~ms} \times 2$).
To find possible overlap, we need to check with the anchor points from connection~2 that are close to 120~ms, \ie 130~ms~($10\mathrm{~ms} + 120\mathrm{~ms}$).
In \autoref{fig:fig04}, point 2~($20 = 120 - 50 \times 2$ and $30 = 130 - 50 \times 2$) is the representative of 120~ms and 130~ms.
By repeating such steps, all the combinations of anchor points can be found and represented by points on the vertical line $x_1 = 20~ms$ in~\autoref{fig:fig04}.

The points are uniform on the vertical line in~\autoref{fig:fig04} due to the constant interval difference of 70~ms.
The least common multiple~(LCM) of the two connection intervals determines the repetition cycle of their behavior.
Any overlap reoccurs after one LCM period.
As a result, there are only limited points on the vertical line.
In \autoref{fig:fig04}, there are 12 points since the LCM of 50 and 120 is 600, and $600 \div 50 = 12$.

We calculate the collision probability for connection~1 by dividing the number of points in the red area by the total points.
However, this is only the collision probability for a specific pair of initial anchor points (20~ms and 10~ms).
Since the two initial anchor points are random in reality, the initial point~(point 0) can be anywhere on~\autoref{fig:fig04}.
Regardless of the starting point, repeating the process gives a vertical line with uniformly distributed points.
Following this observation, the collision probability is the ratio of the red area to the whole figure.
In other words, it is the same as the overlap probability.
Therefore, $P_{12}$ represents both the overlap and collision probabilities of connection~1.

\begin{figure}
	\centering
	\includegraphics[width=1\linewidth]{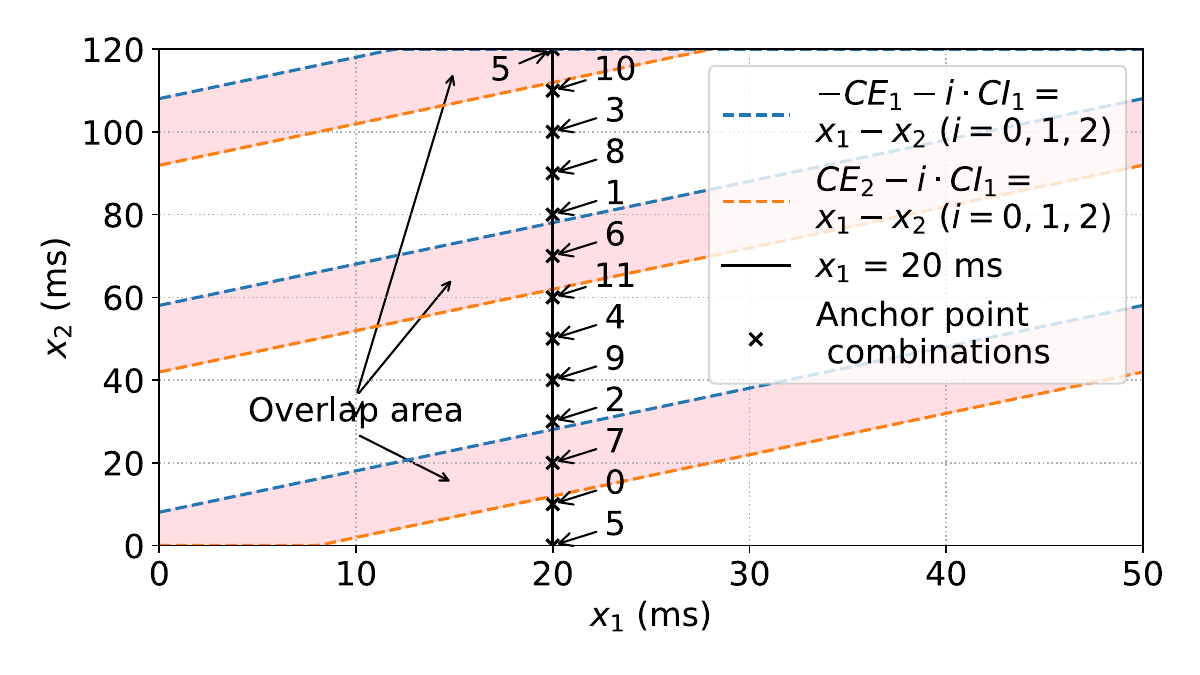}
	\caption{Graphical representation and explanation of \autoref{equation05}. The explanation is the same as \autoref{fig:fig04}, thus not repeated here. But note that this figure represents the case for the connection with a larger interval value.}
	\label{fig:fig05}
\end{figure}

To calculate the collision probability for connection~2, we extend \autoref{fig:fig04} to \autoref{fig:fig05}.
The key difference is the presence of multiple red areas.
The explanation for the expansion of the red area is as follows.
The interval of connection~2 is 120~ms, while 50~ms for connection~1.
So each interval of connection~2 can accommodate 2.4 ($120 \div 50$) intervals of connection~1.
By adding extra red areas, all possible overlaps are considered.
For example, when $x_2 = 60$, there is no overlap in \autoref{fig:fig04}, but if $x_1 = 10$, an overlap occurs at 60~ms.
Therefore, \autoref{fig:fig05} covers all possibilities for connection~2, which \autoref{fig:fig04} does not.

Similarly, we can calculate the collision probability for connection~2 as the ratio between the red area and the whole figure.
To have an approximate result, one can use the Monte Carlo method~\cite{barbu_monte_2020}.
Based on \autoref{equation04} and the explanation above, the mathematical expression of the collision probability for connection~2 is as follows.

\begin{equation}
	\label{equation05}
	\footnotesize
    \begin{aligned}
		P_{21} 
		& = \frac{CI_2}{CI_1} \cdot P_{12} \\
		& = \frac{1}{{CI_1}^2} 
		\int_0^{CI_1} \int_0^{CI_2} 
		1\{-CE_1 < x_1 - x_2 < CE_2\} 
		\,dx_1\,dx_2
	\end{aligned}
\end{equation}

As shown in \autoref{equation05}, one can calculate the collision probability for connection~2 by multiplying a coefficient with \autoref{equation04}.
The coefficient is $\frac{CI_2}{CI_1}$, which represents the number of $CI_1$ accommodated in each $CI_2$.

Up to this point, we have provided the collision probabilities for both connections through \autoref{equation04} and \autoref{equation05}.
To properly investigate the coexistence between BLE connections managed by a single device, both of them are essential.
They represent how frequently a connection collides with the other one.

\subsection{Multiple Connections}
We now extend our mathematical model to cover a more universal scenario, \ie a single BLE device manages more than two BLE connections.
We achieve this by utilizing \autoref{equation04} and \autoref{equation05} between every two connections. 

Consider the collision between connection $i$ and connection $j$, one can calculate $P_{ij}$ and $P_{ji}$, similar to $P_{12}$ and $P_{21}$ in \autoref{equation04} and \autoref{equation05}.
For connection $i$, we use $P_i$ to represent the total collision probability.
It suggests how likely connection $i$ faces at least one collision with all the other connections.

The initialization of any BLE connection highly depends on the process of advertising and scanning, thus, it is random even in a well-designed BLE network. 
Due to the randomness, we consider the establishment of each connection independently.
Therefore, we calculate the collision probability $P_i$ that connection $i$ experiences when the node maintains overall $n$ connections as follows.

\begin{equation}
	\label{equation06}
	\small
	\begin{aligned}
		P_i = 1 - \prod^n_{\substack{j=1, \\ j \neq i}} (1 - P_{ij})
	\end{aligned}
\end{equation}

The term $(1 - P_{ij})$ represents the probability that connection $i$ does not conflict with $j$.

\section{Simplified Mathematical Model for Constrained Deployment}
\label{Model Approximation}
We can clearly explain and accurately calculate the collision probability for both connections based on \autoref{equation04} and \autoref{equation05}, and \autoref{fig:fig04} and \autoref{fig:fig05}.
For engineering use, however, developers prefer a lightweight solution since integral calculation can be computationally intensive on constrained devices.
The Monte Carlo method can already lower the computational overhead, depending on the number of points sampled~\cite{barbu_monte_2020}.
Implementing the Monte Carlo method on IoT devices with constrained resources remains challenging, though.
Furthermore, most embedded systems still use C/C++, and such languages do not support integral calculation well~\cite{glesaaen_c_2024}.
To be able to efficiently approximate the collision probability on constrained IoT devices, we propose the following method.
We emphasize its simplicity with introducing only tiny errors.

\autoref{fig:fig04} represents the collision probability for connection~1.
As shown in \autoref{equation02}, the overlap area highly depends on the lengths of two connection events, \ie $CE_1$ and $CE_2$.
\autoref{equation02} gives the upper and lower limits of the overlap area, \ie $-CE_1 = x_1 - x_2$ and $CE_2 = x_1 - x_2$.
They are parallel to each other.
The vertical distance between these two lines is $CE_1 + CE_2$.
Hence, by extending the lower limit line $CE_2 = x_1 - x_2$ towards the $y$-axis, it will intersect with the $y$-axis and form a parallelogram.
The area of this parallelogram can be calculated as the product of $CE_1 + CE_2$ and $CI_1$---which does not introduce relevant processing overhead.
As a result, we approximate the collision probability for connection~1 as follows.

\begin{equation}
	\label{equation07}
	\small
	\begin{aligned}
		P_{12} 
		& \approx \frac{(CE_1 + CE_2) \cdot CI_1}{CI_1 \cdot CI_2} \\
		& = \frac{CE_1 + CE_2}{CI_2}
	\end{aligned}
\end{equation}

\autoref{equation07} approximates the collision probability for connection~1 by a ratio.
A similar expression can be found in~\cite{pang_novel_2023, pang_modeling_2024}, but it studied the electromagnetic interference between two BLE connections.
Note that the connection event lengths can impact the accuracy of the approximation.
Hence, it depends on the use case if the approximation should be used or not. 
For example, in \autoref{fig:fig04}, both connection events exhibit a length of 8.056~ms, and the connection intervals are 50~ms and 120~ms, respectively.
Based on \autoref{equation04}, the collision probability for connection~1 is 0.1289, while the approximation offers a result of 0.1343.
The error is only 0.0054, thus negligible.

Following the logic between \autoref{equation04} and \autoref{equation05}, we give the approximation of the collision probability for connection~2 as below.

\begin{equation}
	\label{equation08}
	\small
	\begin{aligned}
		P_{21} 
		& \approx \frac{CI_2}{CI_1} \cdot P_{12} \\
		& = \frac{CE_1 + CE_2}{CI_1}
	\end{aligned}
\end{equation}

The only difference is the denominator; it changes from $CI_2$ to $CI_1$.
With the same connection events~(8.056~ms) and intervals~(50~ms and 120~ms), the collision probability for connection~2 and its approximation are 0.3093 and 0.3222, respectively.
The error is 0.0129 which is the product of 0.0054 and $\frac{CI_2}{CI_1}$.
Comparing the collision probabilities (0.1289 and 0.3093), we emphasize the necessity of calculating the collision probability for both connections.
This also partially illustrates the difficulty of setting up BLE parameters properly for a multi-hop context.
The big difference between two connection interval values can lead to a large collision probability for the connection with a larger connection interval, thus an unstable network.

To use the approximation when there are more than two connections, we follow the logic of \autoref{equation06}.
We replace $P_{ij}$ with its approximation.
As a result, we approximate the collision probability $P_i$ as below.

\begin{equation}
	\label{equation09}
	\small
	\begin{aligned}
		P_i = 1 - \prod^n_{\substack{j=1, \\ j \neq i}} (1 - \frac{CE_i + CE_j}{CI_j})
	\end{aligned}
\end{equation}

Till now, we have analyzed the collision probabilities for connections managed by a single device, including their approximation.
However, we still need to involve the impact of clock drift into the mathematical model.
Due to multiple environmental factors, such as temperature and moisture, it is challenging to predict the clock drift~\cite{bilekdemir_clock_2021}.
Regardless of the clock drift characteristics, over the long term, the combinations of anchor points will always be uniformly distributed along the vertical line in the figure.
As a result, the collision probabilities are the same with or without involving the clock drift.
This is later validated in \autoref{Verification Experiment}.

\section{Analysis of Clock Drift and Impact in Practice}
\label{Analysis Experiment}
Before validating our mathematical model, we use practical experiments to illustrate the impact of clock drift in more detail.

\subsection{Experiment Setup}
To give a systematic understanding of the clock drift and its impact, we conduct the following three experiments.

\paragraph{Experiment~(1)}
In this experiment, we follow the minimum setup of a BLE multi-hop network, \ie three BLE devices. 
There are two connections established between the three devices.
One BLE device serves as the central of the other two BLE devices.
In this case, although there are two connections on the central device, we expect no relative drift, thus no collision or impact.
This is a control experiment for the later experiments.

\paragraph{Experiment~(2)}
There are two sub-experiments in this experiment.
For both sub-experiments, we follow the minimum setup of a BLE multi-hop network, \ie three BLE devices.
Again, there are two connections established between the three devices.
In the first sub-experiment, two BLE devices are the central of the third device.
The relative drift between the two central schedules will eventually lead to collision~(see \autoref{Background}).
In the second sub-experiment, one BLE device acts as the central for another BLE device, which then serves as the central for the last device.
As a result, the relative drift should still occur, and eventually collision and performance drop.
Although the roles of each node and the connections in between are different in the two sub-experiments, we expect collision and performance drop to occur in both of them.
This is also to show that changing connection roles or topologies does not mitigate or eliminate the impact of clock drift.

\paragraph{Experiment~(3)}
In this experiment, we extend the first sub-experiment of experiment~(2) from three BLE devices to five.
Four BLE devices are the central of the fifth device~(the only peripheral to all the other devices).
As a result, more connections exist on the peripheral and more collision should occur.
This experiment is to show that more connections on a single BLE device or a more complicated topology can only increase the collision and the performance drop.

\paragraph{Systems configuration}
All three experiments share the following BLE settings.
All the connections use 1M PHY as their physical mode.
All the connections share the same connection interval value of 100~ms.
Each side of a BLE connection sends one packet during every connection interval.
Each packet has a size of 233~bytes, with 200-byte user data inside.
With two packets and two inter frame spaces~\cite{bluetooth_sig_bluetooth_2023}, each connection event lasts for 4.028~ms.
Other BLE settings do not impact the experimental results, such as transmission power, which is why we use the default values.
All experiments are based on the IoT operating system RIOT~\cite{bghkl-rosos-18} and run on Adafruit Feather nRF52840 Sense boards~\cite{industries_adafruit_2024}.

To observe the collision and its impact, we pay attention to three parameters.
\one The time difference between anchor points from different connections, \two the retransmission on the device, and \three the disconnection occurred on each device.
Since all the connection interval values are the same, the time difference between anchor points directly shows the relative position and drift between the connections.
The collision between connection events should occur when the following condition is met.
The time difference is close to or smaller than the length of a single connection event~(4.028~ms).
Once the collision occurs, the BLE device should experience a dense retransmission period because the blocked connection events prevent the packets from being sent.
With collision or retransmission, we expect to see disconnection and reconnection on the devices.

\subsection{Results}
\begin{figure*}[]
	\centering
	\begin{subfigure}[t]{0.24\textwidth}
		\includegraphics[width=\linewidth]{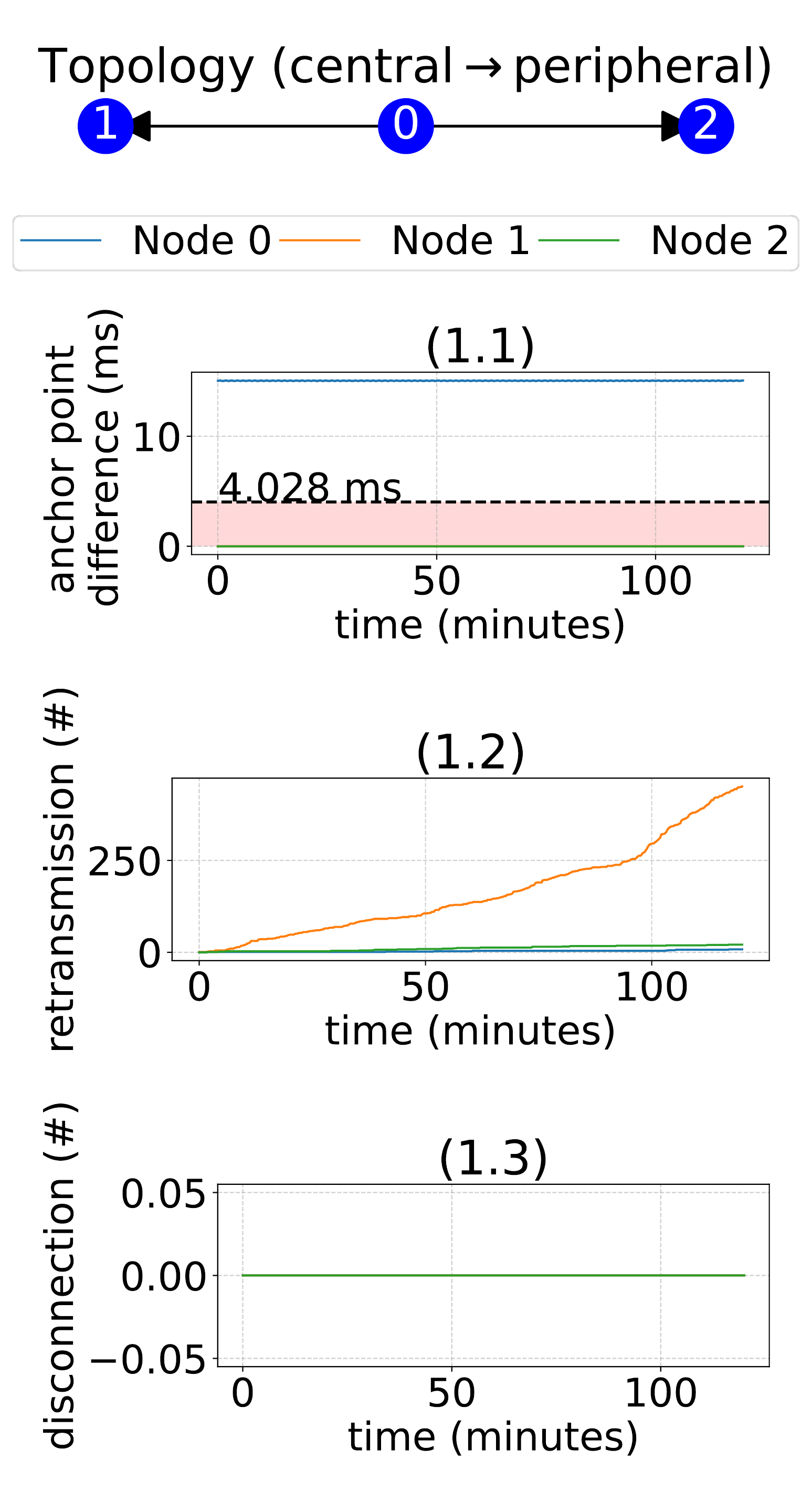}
		\caption{Experiment~(1). One central node connects to and controls two peripheral nodes, thus no clock drift shows on the central node.}
		\label{fig:fig06_01}
	\end{subfigure}
	\hfill
	\begin{subfigure}[t]{0.24\textwidth}
		\includegraphics[width=\linewidth]{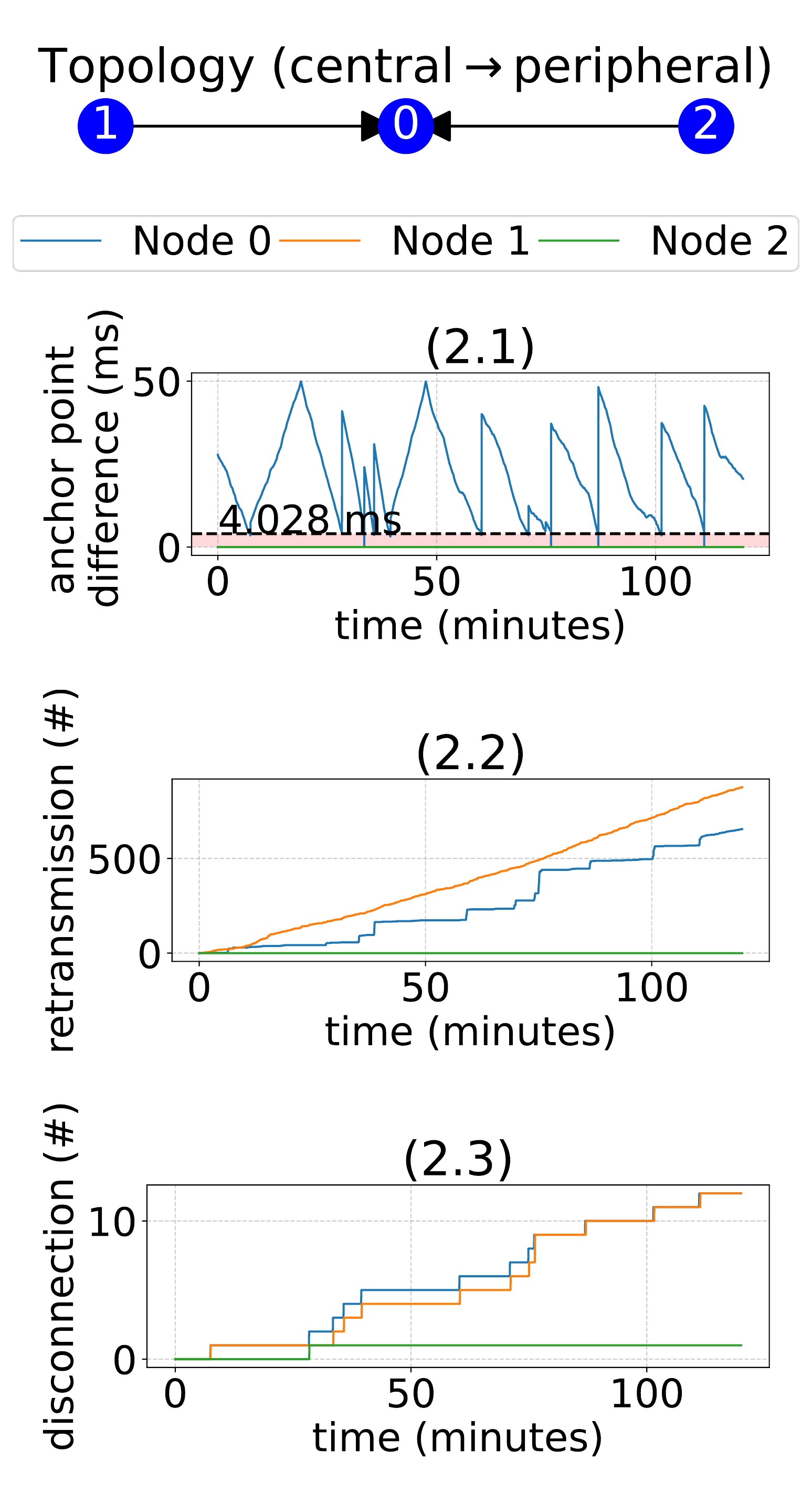}
		\caption{First sub-experiment of experiment~(2). Two central nodes connect to and control one peripheral node, thus the impact of clock drift shows on the peripheral node.}
		\label{fig:fig06_02}
	\end{subfigure}
	\hfill
	\begin{subfigure}[t]{0.24\textwidth}
		\includegraphics[width=\linewidth]{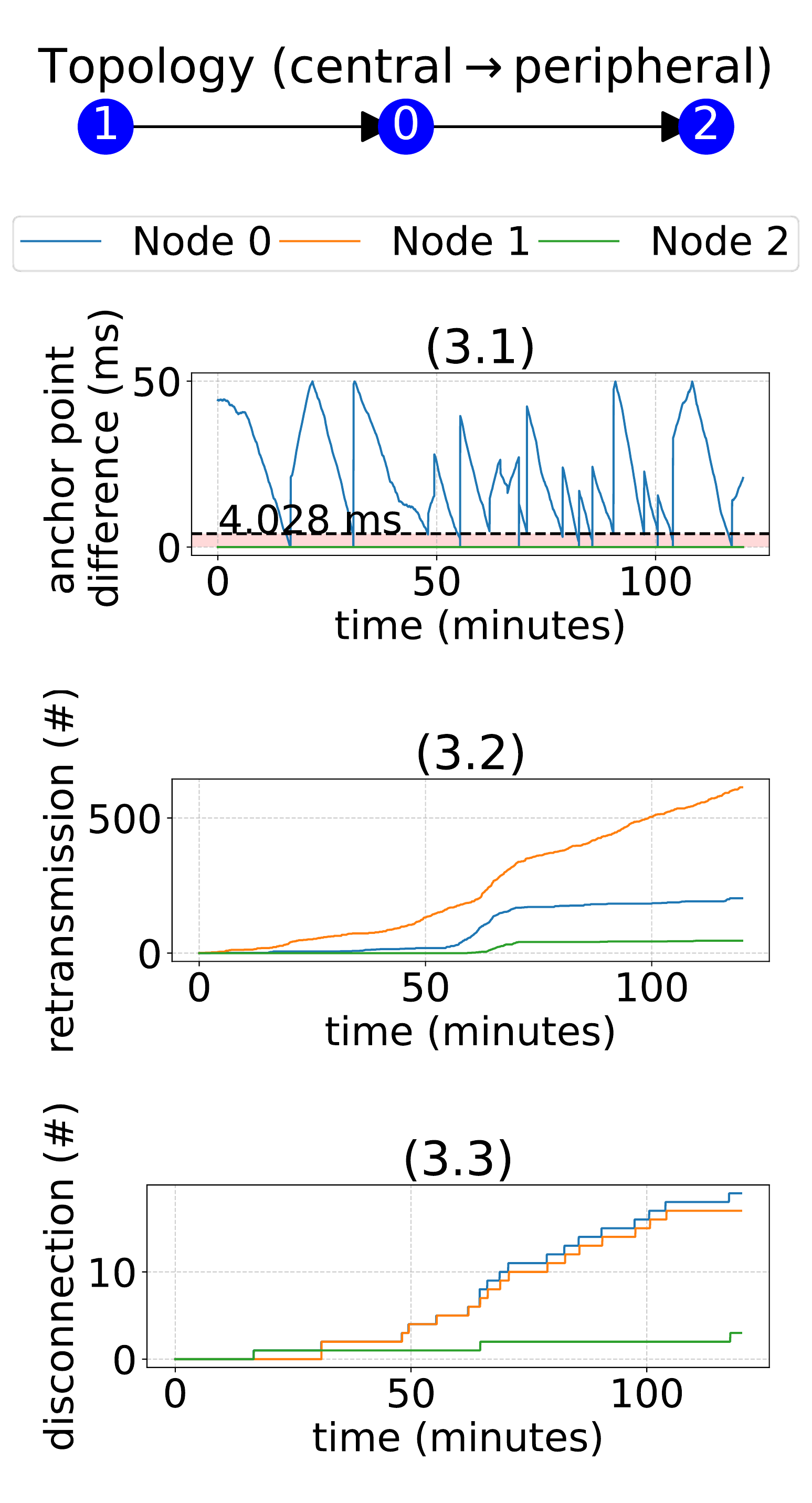}
		\caption{Second sub-experiment of experiment~(2). Node~0 connects as the central to node~2, and connects as the peripheral to node~1. The impact of clock drift also show.}
		\label{fig:fig06_03}
	\end{subfigure}
	\hfill
	\begin{subfigure}[t]{0.24\textwidth}
		\includegraphics[width=\linewidth]{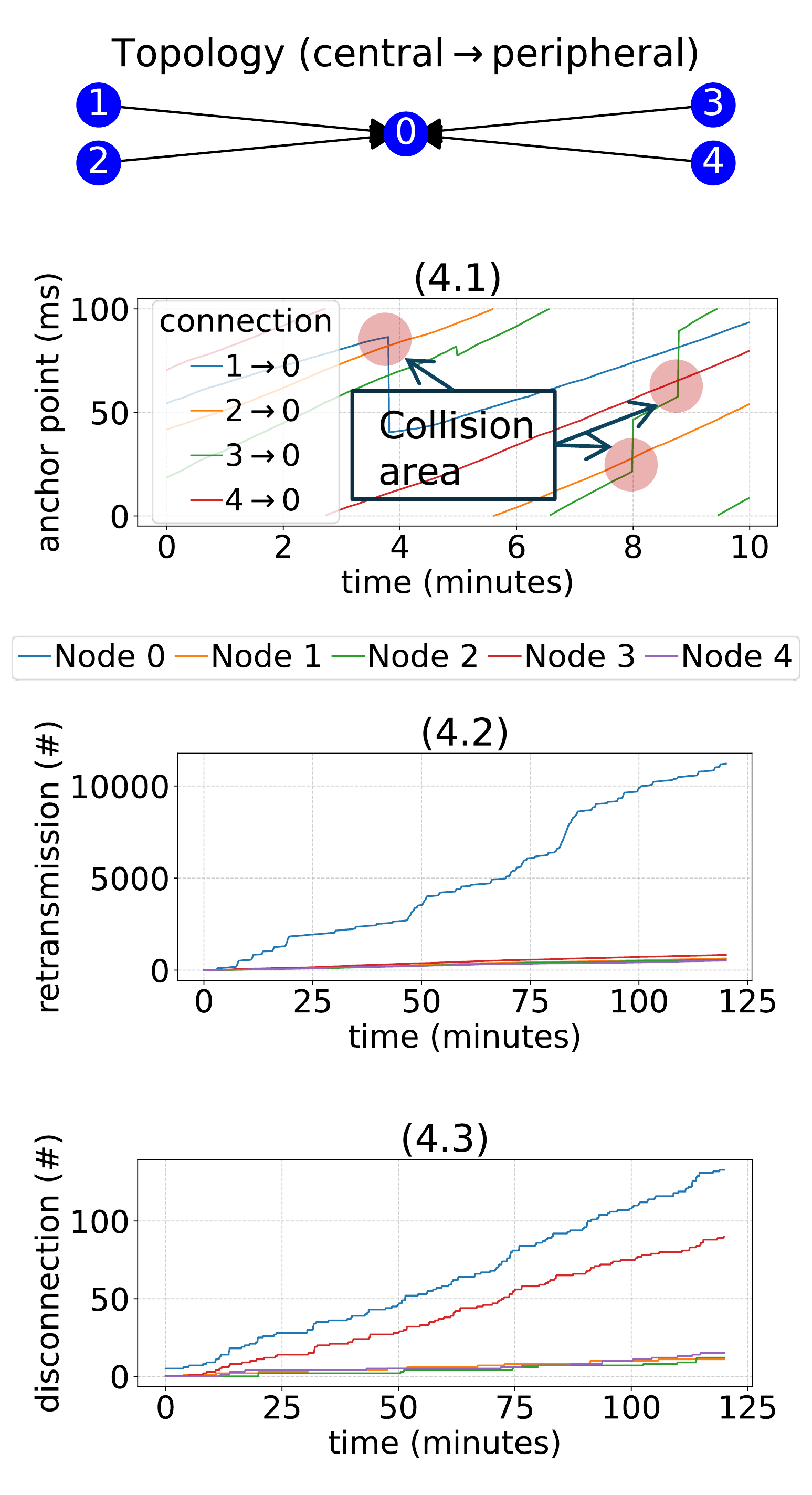}
		\caption{Experiment~(3). One node connects as the peripheral to all the other four nodes. Connection collision occurs more frequently, thus more performance drop and connection loss.}
		\label{fig:fig06_04}
	\end{subfigure}
	\caption{Experiment results of multi-hop BLE to show the impact of clock drift.}
	\label{fig:fig06}
\end{figure*}

\autoref{fig:fig06} includes all the results from the experiments.
Each column represents a setup, and each row describes a parameter.

\paragraph{Experiment~(1)}
\autoref{fig:fig06_01} illustrates the results of experiment~(1).
We show the topology between the three nodes on the top of the graph, \ie a typical star topology from one central device~(node~0) to two peripheral devices~(node~1 and node~2).
In the graph, we can see that the difference between the two connections managed by the central node stays constant.
It means that there is no relative drift between the two connections.
The difference between the two neighboring anchor points on the central device is constant to be around 15~ms.
Considering the connection interval of 100~ms and the connection event of 4.028~ms, the 15~ms should not cause any collisions.
The packet retransmission shows quite low on node~0 and node~2, but rather high on node~1.
There might be two reasons.
First, the interference in the air may impact some BLE packets.
Second, the BLE stack may have some self-conflict when managing two or more connections on a single device, and shows such impact specifically on one or some random devices.
The disconnection graph shows that there is no disconnection happening during the 2-h experiment.

\paragraph{Experiment~(2)}
In the first sub-experiment of experiment~(2) (\autoref{fig:fig06_02}), we reverse the role of each device in experiment~(1) to one peripheral device~(node~0) plus two central devices~(node~1 and node~2).
As expected, the clock drift exists on the peripheral device, shown as the difference between the anchor points of the two connections managed by node~0 varies all the time.
The difference value changes between 0~ms and 50~ms, since in this experiment we monitor the closest distance between the anchor points of the two connections.
Therefore, the maximum value is half of the connection interval.
Due to the clock difference between the two central nodes, connections relatively move on the peripheral node, and the impact shows.
In the retransmission graph, we can see that node~0 has its retransmission in a stepped feature, which suggests fast increase on the retransmission.
Such periods match perfectly with the time when the anchor point difference reaches and drops below 4.028~ms, \ie when two connection events start to collide on the peripheral node.
A similar phenomenon shows in the disconnection graph as well.
Around the time when the anchor point difference reaches or drops below 4.028~ms, the disconnection and reconnection occur.
They are the typical negative impact of clock drift on a BLE multi-hop network, \ie performance drop and connection loss.

Although with a different topology~(see~\autoref{fig:fig06_03}), the second sub-experiment of experiment~(2) shows quite similar results and features to the first sub-experiment.
It suggests that as long as a device does not play the central role in all of its connections, the impact of clock drift shows.
Due to the similarity of the results, we do not repeat the discussion.

\paragraph{Experiment~(3)}
In the last column of~\autoref{fig:fig06}, we show the experimental results when there are four connections existing on a BLE node.
This is to show that the impact of clock drift only becomes worse when there are more connections on each BLE device.
In~\autoref{fig:fig06_04}, we plot the drifts of the four connections on node~0, but only for the first 10~min, since the 2-h results are too crowded and cannot show details.
Although all the connections drift with the same direction, but their drift speeds are slightly different and are not constant.
This is the reason why the vertical distances among the curves gradually change.
When the vertical distances are close to or smaller than 4.028~ms, there is a collision between two connections, such as the collision area marked in the graph.
Similar to the phenomenon from previous experiments, when a collision occurs, there is disconnection and reconnection.
As a result, we see the sudden change of the anchor points on the BLE node~0.
The reconnection locates the anchor point at a random time point on node~0.
However, due to more connections on node~0, the disconnection and reconnection are more frequent.
We can easily see this by comparing the disconnection number with previous experiments.
The performance drop shows in the dramatic increase in the retransmission.

\section{Verification of Mathematical Model}
\label{Verification Experiment}
According to the phenomena in~\autoref{Analysis Experiment}, we verify our mathematical model.
The verification is also through practical experiments on Adafruit Feather nRF52840 Sense and RIOT.

\subsection{Experiment Setup}
We implement the minimum use case of a BLE multi-hop network, \ie two connections managed by a single BLE device, similar to our analysis experiments~(see~\autoref{Analysis Experiment}).
According to~\autoref{Analysis Experiment}, when a BLE node is the central to the other two, there is no impact of clock drift shown.
Therefore, we only design our verification experiments on the other two possible topologies.
They are the same as experiment~(2) in~\autoref{Analysis Experiment}.
Despite of different topologies, they should both provide results that match with our mathematical model.
To enhance the validation, we also choose different connection parameters, such as connection intervals and packet number in each connection interval.
This is also because the case when both connections share the same parameters has been shown and discussed in~\autoref{Analysis Experiment}.
In general, we use the following two scenarios for the validation.

\paragraph{Scenario~(1)}
In this scenario, we ask two devices to be the central of the third device.
We design the two connections with different connection intervals.
The first connection has a fixed interval of 50~ms.
The second one varies its connection interval from 60~ms to 260~ms, with an increment of 50~ms each time.
Both connections have a connection event of 4.028~ms, \ie a 233-byte packet from both the central and the peripheral side.
We measure the overlap probability for all the combinations and compare with results from both the theory and approximation in~\autoref{Mathematical Model} and \autoref{Model Approximation}.

\paragraph{Scenario~(2)}
In this scenario, one BLE node acts as the central for another BLE device, which then serves as the central for the last node.
One of the connections has a fixed interval of 100~ms.
The other connection changes its interval from 105~ms to 125~ms, with an increment of 5~ms each time.
We design the connection events in each connection to be 8.056~ms, \ie two 233-byte packets from both the central and the peripheral side.
We measure the overlap probability for all the combinations and compare them with the theory and approximation.

Since we cannot control the connection establishment procedure, the initial anchor points of the two connections randomly locate on the device which manages two connections.
Due to different connection interval values, we do not monitor the relative movement between every two anchor points from the two connections anymore.
Instead, we use the LCM mentioned in~\autoref{Mathematical Model} to observe the relative movement.
For instance, with the interval combination of 50~ms and 60~ms, we monitor the 50-ms connection every 6 connection intervals, and the 60-ms connection every 5 connection intervals.
This way, we can observe the relative movement between the two connections on the BLE node.

\subsection{Results}
\begin{figure}
	\centering
	\includegraphics[width=1\linewidth]{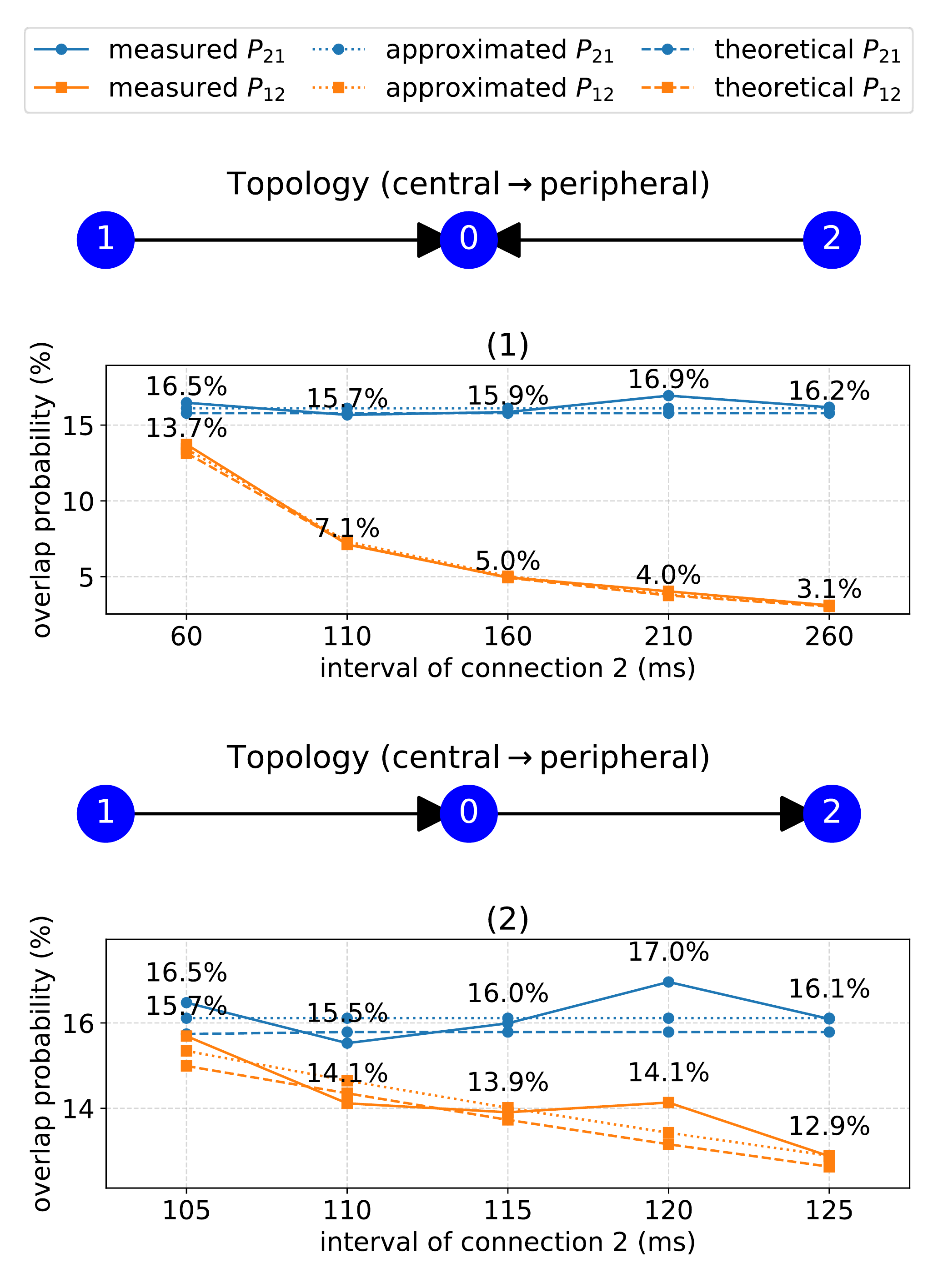}
	\caption{The comparison among the measurement, approximation, and theory. Although with different connection parameters, the similarity shows in both scenarios.}
	\label{fig:fig07}
\end{figure}

We report the measurement in \autoref{fig:fig07}, and compare to the theoretical and approximated results from the model in \autoref{Mathematical Model}.
The figure shows three results for each connection, \ie the measured one, the approximated one, and the theoretical one.

\paragraph{Scenario~(1)}
In~\autoref{fig:fig07}(1), the similarity among all the three results is evident.
Such a similarity illustrates the accuracy of our mathematical model in the large scale, \eg when varying the connection interval from 60~ms to 260~ms.
There are some deviations between the measured results and the other two calculated results, but it is normal that the practice does not match with the theory perfectly.
We observe that such deviations decrease with the increase of the measurement time, and the measurement eventually converges to the theory.
The two overlap probabilities~($P_{21}$ and $P_{12}$) show different trends.
We calculate $P_{21}$ as a horizontal line, and the measurement follows the trend.
This is because $P_{21}$ represents the collision probability for the connection with a larger interval, \ie 60~ms to 260~ms, and it is only related to the smaller connection interval according to~\autoref{equation05}.
However, the collision probability for the connection with a smaller interval, \ie 50~ms, decreases since the other connection increases its interval.
This is easy-understanding since increasing the connection interval reduces the interference probability for the connection with a fixed interval.
But the connection that changes its interval is under the same level of interference all the time.
The maximum error shows on $P_{21}$ when the interval of connection~2 equals to 210~ms, which is around 1\% only.

\paragraph{Scenario~(2)}
\autoref{fig:fig07}(2) gives the results of scenario~(2).
This experiment gives a comparison in a small scale, \ie when varying a connection interval from 105~ms to 125~ms.
Again, the similarity between the measured results and the two calculated results validates the mathematical model.
Although with different connection parameters, the results are quite similar to the ones reported in~\autoref{fig:fig07}(1).
Due to the larger connection interval, it needs more time for the measurement to converge to the theory.
For instance, in~\autoref{fig:fig07}(1) it only takes around 1~h to have the convergence.
But in~\autoref{fig:fig07}(2), the measurement needs around 3~h to converge to the theory.
For the same reason, the two overlap probabilities show different trends, thus we do not repeat the explanation.
$P_{21}$ curve has the largest error between the measurement and the theory, which is around 1\% again.

\section{Related Work}
\label{Related Works}
\paragraph{Clock drifts in BLE multi-hop networks}
To the best of our knowledge, Petersen \etal \cite{petersen_mind_2021} are the first to report on the impact of clock drift in BLE multi-hop networks.
To mitigate the problem, they proposed the usage of random connection intervals instead of the same value.
All evaluation is empirical.
In this paper, we close the gap by introducing a mathematical model to cover general deployment cases and derive a systematic understanding of limits and potentials of multi-hop BLE networks.

\paragraph{Time synchronization}
Time synchronization and calibration have been challenging for wireless sensor networks for years.
\citeauthor{romer_time_2001}~\cite{romer_time_2001} first discussed these challenges in the context of  ad hoc networks.
He presented a time synchronization scheme that was suitable for sparse ad hoc networks.
However, due to the difference between ad hoc network access and BLE, the scheme cannot be applied in BLE multi-hop networks. 
\citeauthor{elson_wireless_2003}~\cite{elson_wireless_2003} found that time synchronization schemes developed for traditional networks such as NTP are not suitable for wireless sensor networks.
They also suggested more appropriate approaches, such as using multiple and tunable modes of synchronization and not maintaining a global timescale for the entire network~\cite{elson_wireless_2003}.
\citeauthor{romer_time_2005}~\cite{romer_time_2005} discussed common approaches for synchronization, their underlying models, synchronization techniques, and algorithms.
They also presented common techniques for evaluating synchronization algorithms and selected evaluation results~\cite{romer_time_2005}.
However, none of them can help with the time synchronization issue within BLE multi-hop networks.

\paragraph{Mathematical models}
The collision between two BLE connections can be modeled as the periodic pulse train/signal problem~\cite{grami_chapter_2016}, a classic mathematical question that tries to answer when two periodic pulse trains/signals meet~\cite{grami_chapter_2016}.
We review related studies, but none of them answers our research questions.
Moreover, our graphic method is more comprehensible.
Below, we discuss these studies.

\citeauthor{richards_probability_1948}~\cite{richards_probability_1948} studied this problem first.
He focused on the probability of a given minimum duration of overlap.
The proposed solution conflicts with practical application, as noted in the study.
Ten years later, \citeauthor{stein_statistical_1958}~\cite{stein_statistical_1958} studied the problem from the perspective of random pulse trains.
They formulated the problem statistically to describe the time coincidences among a set of random pulse trains.
Later, \citeauthor{hatcher_ew_1976}~\cite{hatcher_ew_1976} proposed solutions to the problems of intercept probability and time of intercept, specifically for surveillance and reconnaissance systems.

After years, the research on the periodic pulse train issue carried on.
\citeauthor{self_intercept_1985}~\cite{self_intercept_1985} continued research on the periodic pulse train by thoroughly investigating the intercept time between two periodic pulse trains, with the background of electronic warfare surveillance and reconnaissance.
Their formulas gave a representative basis for the calculation of intercept time statistics.
\citeauthor{kelly_synchronization_1996}~\cite{kelly_synchronization_1996} studied the probability of coincidence between two pairs of pulse trains.
\citeauthor{clarkson_numbertheoretic_1996}~\cite{clarkson_numbertheoretic_1996} investigated a number of problems concerning the overlaps or coincidences of two periodic pulse trains.

Most recently, \citeauthor{jiang_modeling_2020}~\cite{jiang_modeling_2020} modeled the probability of partial pulse overlap between two periodic pulse signals.
The research calculated the overlap probability of part of one pulse and part of pulse-group between two periodic pulse signals.
The authors validated their models through simulation, thus they can use the model to analyze interference between two radio waves.
The same researchers further analyzed the overlap in the time domain between two periodic pulse signals~\cite{jiang_analyzing_2021}.
They focused on the characteristics of overlap during an arbitrary observation time.
They introduced an analytical model to calculate the average probability of overlap based on the joint probability distribution function.
The research results were considered helpful to analyze electronic interference and provide methods on anti-jamming.

The limitation of using BLE in multi-hop networks is a typical periodic pulse train/signal problem.
However, none of the existing publications answers the research question of this paper.
Besides, our paper proposes the graphic method to dramatically simplify the mathematical procedure and its understanding.
Meanwhile, we verify our model through practical experiments, and shows that the simplicity does not impact its accuracy.

\section{Discussion}
\label{Limitation_Analysis}
The collision of a BLE multi-hop network depends on three factors: connection events, connection intervals, and the number of connections on BLE devices.
In most cases, the BLE application and network determine all the three factors.
Hence, once the application and network are fixed, the collision probabilities are also fixed for all the connections within the network.
If the application and network are flexible, the three possible ways to minimize the collision of the multi-hop network are decreasing the lengths of connection events, reducing the number of connections on each BLE device, and increasing the connection interval values.
All three options, however, have drawbacks.
Decreasing the length of connection events would lead to a low throughput.
Reducing the number of connections lowers the scalability and flexibility of the network.
Increasing the connection interval values negatively impacts the latency.
Hence, we do not consider them a proper solution.
We further discuss the possibility of setting up a proper BLE multi-hop network by answering the following two research questions:
\emph{Can a BLE multi-hop network avoid collision?}
and
\emph{Can a BLE multi-hop network manage its collision frequency?}

\subsection{Can a BLE Multi-hop Network Avoid Collision?}
To allow a BLE device to manage multiple connections without any collisions, the only topology is the star topology.
Moreover, the device that manages multiple BLE connections must be the central.
This way, all the connections follow the same schedule, and zero relative drifts can be ensured among the multiple connection schedules.

According to BLE specification, a central can adjust the anchor points for the peripherals connected to it.
As a result, for a more complicated topology, such as line and tree, the possibility of having no or little collision still exists.
However, it needs certain steps and mechanisms to achieve so.
Firstly, each BLE device should only play the peripheral role in maximum one connection.
Except for this connection, the device should be the central for all the other connections.
Secondly, there is a need for a mechanism to monitor and adjust the anchor points.
These two steps help reduce the collisions among BLE connections, by changing connection schedules when they are about to collide.
Note that, without such steps and mechanisms, a topology of line and tree will for sure introduce collision (see the results in \autoref{Analysis Experiment}).

Without changing BLE standard or hardware, it is impossible for a mesh or fully-connected BLE multi-hop network to avoid collisions.
This is because each device may hold the peripheral role several times, and BLE standard does not allow a peripheral device to adjust the connection schedules or anchor points.
We propose two possible directions to mitigate the clock drift issue if a BLE multi-hop network must be established in mesh or fully-connected topology.
However, they need some changes either on software or on hardware.

\paragraph{Possible directions by changing BLE standard}
We suggest introducing a strategy for BLE devices, no matter as a central or a peripheral, to have full control on the anchor points.
This strategy can fill in the gap that a peripheral cannot adjust the anchor points for the centrals connected to it.
Once with the strategy, the topology of a line or tree will not be a limitation anymore.

\paragraph{Possible directions by changing hardware}
The essence of collision is actually an issue of hardware occupation.
Hence, adding extra hardware, i.e., transceivers, should solve the problem.
This way, we can transfer the hardware occupation issue to an interference challenge between BLE connections.
Although the interference can also lead to reliability drop~\cite{pang_novel_2023}, it should be much less compared with the drop due to collision.
Besides, the development of full-duplex wireless communication~\cite{jain_practical_2011} should help further improve the reliability.
This idea is similar to the Multiple Input Multiple Output (MIMO) in Wi-Fi~\cite{he_wifi_2020}.

\subsection{Can a BLE Multi-hop Network Manage Its Collision Frequency?}
Considering long-term deployments with static connection setups, the frequency of collisions is fixed (see \autoref{fig:fig04} and \autoref{fig:fig05}).
However, it does not suggest that connection parameters should be randomly set.
BLE designers and developers should apply certain constraints to the parameters, because such constraints can prevent the BLE connections from short-term continuous collisions or long-term high-frequency collisions.
We identify three potential constraints.

\paragraph{The connection interval values should not be too close to each other}
Instead, the determination of them should consider the lengths of connection events.
Imagine two connection interval values are close to each other, e.g., only 1.25~ms different, meanwhile the connection events are long, like 10~ms.
Once the two connections start to collide, short-term but continuous collisions begin.
This is because the difference of 1.25~ms is too small comparing with the large connection events.
As a result, the connections cannot get through the collision area fast, which suggests the collision occurs between the two connections continuously, for like 10 to 20 times.
Although a short period, it can be fatal to the connections.

\paragraph{The LCM between any two connection intervals should not be too small}
The optimal connection interval values can be mutually prime numbers.
As an example, two connection intervals~(80~ms and 120~ms) are far from each other, but still impracticable.
The LCM between 120 and 80 is 240.
It suggests that once the collision occurs between these two connections, it will repeat every two connection intervals for the 120-ms connection and every three connection intervals for the 80-ms connection.
Moreover, the two connections have to undertake the collisions for long to get through the collision period due to the small clock drift.
With such a high collision frequency for a long time, the two connections are not reliable or stable anymore.
By setting the two connection intervals to mutually prime numbers or at least ensuring a large LCM between them, BLE connections can avoid the long-term high-frequency collisions.

\paragraph{The large difference between two connection intervals may result in too many collisions on the connection with the larger connection interval}
\autoref{fig:fig05} illustrates this issue already.
The large difference will increase the number of overlap areas in \autoref{fig:fig05}, thus a high collision probability for connection~2.
Combining with the first constraint, connection intervals on a single BLE device must be within a certain range.
The lower limit is related to connection event lengths (to prevent continuous overlaps), while the upper limit is influenced by the collision probability (as in \autoref{fig:fig05}).
This range ensures the avoidance of short-term continuous collisions and too high collision probabilities for connections with large intervals.

In general, the current BLE time synchronization mechanism cannot handle the clock drift challenge in multi-hop contexts.
The above constraints limit the number of connections that a BLE device can properly manage.
However, by knowing the above constraints, the designers and developers of BLE multi-hop networks can at least determine if some connection settings are feasible or not.

\section{Conclusions and Future Work}
\label{Conclusions}
This research is the first one to reveal the impact of clock drift on BLE multi-hop networks.
We develop a mathematical model to analyze collisions between BLE connections managed by a single device.
It accurately calculates collision probabilities and visualizes the process via a graphical method.
Results show that current BLE standards are unsuitable for efficient multi-hop networks.
Fine-tuning parameters may help avoid short-term issues but remain insufficient.
To address the collision issue, we propose: (1) limiting the topology of BLE multi-hop networks; (2) giving BLE devices full control on the anchor points or adding a mechanism of encoder and decoder; and (3) adding transceivers to transform hardware constraints into manageable electromagnetic interference.
We consider this research a cornerstone for future development of BLE multi-hop networks.

As for future work, we aim to promote BLE multi-hop networks by implementing our proposed solutions, e.g., modifying the BLE protocol and adding extra transceivers.
Additionally, we would also like to evaluate the performance of a BLE multi-hop network under Wi-Fi interference to enable practical deployment.

\section*{Acknowledgments}
\label{Acknowledgments}
This work was supported in parts by the German Federal Ministry of Research, Technology and Space (BMFTR) within the research project C-ray4edge (grants 16KIS1694K and 16KIS1695).

{\footnotesize
	\bibliographystyle{IEEEtranN}
	\bibliography{IEEE_IoTJournal_03}
}

\newpage

\vfill
\end{document}